\documentclass[12pt]{article}
\usepackage{amsmath,amssymb,latexsym,float}
\usepackage{enumerate}

\numberwithin{equation}{section}

\newcommand{\W}{\mathcal{W}}

\newcommand{\R}{\text{\fontshape{n}\selectfont I\kern-.42exR}}

\newcommand{\1}{\text{\fontshape{n}\selectfont 1\kern-.56exl}}

\begin{document}
\title{
{\normalsize \hspace{10cm}CERN-PH-TH/2010-246}\\
{\bf Weyl and ghost fermions on the lattice}
}

\author{Artan Bori\c{c}i\\
        {\normalsize\it CERN, Physics Department, Theory Division, 1211 Geneva 23, Switzerland}\\
        {\normalsize\it and}\\
        {\normalsize\it University of Tirana, Physics Department, Zog I Bvd, Tirana, Albania}\footnote{Permanent address}\\
}

\date{}
\maketitle

\vspace{2cm}
\begin{abstract}
Nielsen-Ninomiya theorem forbids Weyl fermions on the lattice which respect the full hypercubic symmetry. By giving up this assumption in a specific way, it is possible to formulate a lattice theory with a single Weyl fermion in four dimensions and a sextet of Dirac particles in two dimensions. This way, the meaning of the theorem in relation to the doubling problem on the lattice is clarified. Whether the proposal will be suited for future lattice computations will depend on the effects the extra particles will have in the interacting theory.
\end{abstract}

\vspace{14cm}
\pagebreak


\setcounter{section}{1}

Formulating Weyl fermions on the lattice is a grand challenge: twenty nine years ago Nielsen and Ninomiya have proved that it is impossible to formulate a local Weyl fermion on a hypercubic lattice \cite{NN_1981}. The assumptions of the theorem restrict the momentum space operator such as it has to intersect the Brillouin zone an even number of times. At each intersection we get fermion theories which differ only by the signs in front of the momenta. This is the way the doublers proliferate. Later it was shown that giving up the hypercubic symmetry one can decrease the number of doublers to a minimum \cite{minimal_doubling}. In this letter we show that the hypercubic symmetry may be broken in such a way that the extra zeros yield particles without any physical relevance and can be treated as artifacts of the lattice regularised theory.

The letter is organised as follows: first, we define the new fermion operator and show that it describes eight fermion degrees of freedom in total. Then we show that two such degrees describe a Weyl fermion and the rest of them are six Dirac fermions in two dimensions. We end the letter with some specific remarks and conclusions.

{\bf 1}. Our momentum space operator reads:\footnote{Basic definitions and unexplained notations are summarised in Appendix A.}
\begin{equation}
\label{weyl_latt}
a\W(p):=e^{iap_4}-1 +i\sum_k~\left(e^{i\sigma_kap_k}-1\right)\ ,
\end{equation}
where $a$ is the lattice spacing and $\sigma_k,k=1,2,3$ are the Pauli matrices. Its classical continuum limit is given by:
$$
\W(p)\rightarrow ip_4-\sum_k~\sigma_kp_k\ ,
$$
yielding thus a Weyl fermion as expected. From now on, for the ease of writing, we will drop the lattice spacing from the notations and take the continuum limit by formally sending momenta to zero.

It is easy to check that the new operator has the following six zeros:
\begin{eqnarray*}
p_1^\pm=(\pm\frac\pi2,0,0,\frac\pi2)\ ,\\
p_2^\pm=(0,\pm\frac\pi2,0,\frac\pi2)\ ,\\
p_3^\pm=(0,0,\pm\frac\pi2,\frac\pi2)\ .
\end{eqnarray*}
Indeed, at $p_1^\pm$ we have:
$$
\W(p_1^\pm)=i-1+i(\pm~i\sigma_1-1)=-1\mp\sigma_1\ ,
$$
i.e. one zero for each sign at a time since the right hand side is a projector in the spin space. Similarly, one can check the zeros in the other cases. In fact, by solving the secular equation,
\begin{eqnarray*}
0=\det~\W(p)&=&(\cos p_4-1)^2-[\sin p_4+\sum_k(\cos p_k-1)]^2-\sum_k\sin^2p_k\\
&+&2i~(\cos p_4-1)~[\sin p_4+\sum_k(\cos p_k-1)]\ ,
\end{eqnarray*}
one can show that there are no more zeros than those stated already.\footnote{Appendix B describes the details of the procedure for the solution of the secular equation.}

{\bf 2.} It is important to recognise now that the extra zeros located at $p_k^\pm,k=1,2,3$ do not duplicate the Weyl fermion at the origin. This is most easily seen if we write the operator in terms of momenta centered at the extra zeros. For example, at $p_3^\pm$ we have:
\begin{eqnarray*}
\W(p_3^\pm+q)&=&-1\pm\sigma_3\cos q_3+i(\cos q_4-1)+i\sum_{l\neq3}(\cos q_l-1)\\
&-&\sin q_4\mp i\sin q_3-\sum_{l\neq3}\sigma_l\sin q_l\ ,
\end{eqnarray*}
which, if expanded in powers of momenta, yields:
$$
\W(p_3^\pm+q)=-1\pm\sigma_3-q_4\mp iq_3-\sigma_1q_1-\sigma_2q_2+O(q^2)\ .
$$
It is clear that the projector operator $-1\pm\sigma_3$ destroys the structure of a possible Weyl particle in four dimensions: it picks up only one of the signs in front of $q_3$ at a time, such that the dispersion relations of the extra particles are determined by the equations:
\begin{eqnarray*}
iq_4&=&+q_3\ ,~~~~iq_2=\pm q_1\ ,~~~~~~\text{at}~p=p_3^+\\
iq_4&=&-q_3\ ,~~~~iq_2=\pm q_1\ ,~~~~~~\text{at}~p=p_3^-\ .
\end{eqnarray*}
These zeros describe particles with definite helicity in the 3 direction associated by a Dirac particle in the 1-2 plane. Clearly, such zeros are not doublers. However, since the definite helicity states come in pairs with opposite helicity, taken together they describe two Dirac particles restricted in the 1-2 and 3-4 planes. This removes any doubt that such a pair may form a possible Weyl doubler, which should move unrestricted in four dimensions. The same analysis applies for the zeros at $p_1^\pm$ and $p_2^\pm$.

{\bf 3.} The lattice fermion proposed in this letter contains eight zeros: two of them describe a Weyl fermion and the other six describe a sextet of two-dimensional Dirac particles. Therefore, this is the first lattice theory that describes a single Weyl fermion without doublers which is also local in the strict sense. There are some other remarks in order here:

\begin{itemize}
\item Formulation of gauge invariant chiral gauge theories has proven to be a complicated and difficult task \cite{Luescher}. In our case, the introduction of the gauge fields can be done in the gauge invariant way as it is usual on the lattice. Therefore, one can in principle formulate a chiral gauge theory on the lattice which is local and gauge invariant. However, the extra particles that accompany the Weyl fermion may complicate the situation. Although they should not be given any physical relevance except from being treated as lattice regularisation artifacts, they may have the effect that doublers usually have in the cancellation of the chiral anomaly \cite{private_comm_Luescher}. Nonetheless, this feature has to be explicitly shown in a separate calculation.
\item Note that the theory violates the hypercubic symmetry and, as a consequence, the Hamiltonian of the corresponding quantum mechanical theory is not guaranteed to be self-adjoint. Therefore, one has to perform similar investigations as in the case of minimally doubled actions in order to restore the hypercubic symmetry \cite{Capitani_et_al}. However, the presence of the extra particles in the continuum limit may not allow a complete restoration in our case.
\item The Weyl fermion can be used to construct a single Dirac fermion in the usual way:
\begin{equation}
\label{dirac_latt}
D(p):=
\begin{pmatrix}
0 & \W(p)\\
-\W(p)^* & 0
\end{pmatrix}
=\sum_\mu i\gamma_\mu\sin p_\mu + \sum_\mu\delta_\mu(\cos p_\mu-1)\ ,
\end{equation}
where $\gamma_\mu,\mu=1,2,3,4$ are the Dirac gamma-matrices in the chiral representation and
$$
\delta_k=i\gamma_4\ ,k=1,2,3\ ,~~~~\delta_4=\gamma_5\gamma_4\ .
$$
Note that, if the term multiplying $\delta_4$ is missing, we recover the Karsten-Wilczek action, which has eight zeros describing two Dirac fermions in four dimensions \cite{minimal_doubling}. As in the case of the Weyl theory, the Dirac theory comes with extra particles, which should be considered as lattice artifacts. Hence, we have, for the first time, a chiral Dirac theory on the lattice which is undoubled and local.
\end{itemize}
In this letter we have formulated a Weyl fermion on the lattice without doublers in expense of the hypercubic symmetry. This way, we have been able to clarify the meaning of the Nielsen-Ninomiya theorem in relation to the doubling problem on the lattice. Whether the present proposal will be suited for future lattice computations with Weyl or Dirac fermions will depend on the effects the extra particles have in the interacting theory.

\section*{Acknowledgements}
The author would like to thank Martin L\"uscher for comments related to this work and the CERN Theory Division for the kind hospitality while this letter was concluded.

\section{Appendix A: Basic notations and definitions}

A fermion degree of freedom is nothing but a spin 1/2 degree of freedom. As a consequence, a fermion field $\phi(x)$, is the two-component complex valued field located at some point $x$ in the four dimensional Euclidean space. The equation that describes the relativistic dynamics of such a field is the Weyl equation:
$$
(\partial_4\pm i\vec{\sigma}\nabla)\phi(x)=0\ ,
$$
where one should choose between + and - signs. It corresponds to the dispersion relation:
$$
ip_4\mp\vec{\sigma}\vec{p}=0\ .
$$
A Dirac fermion field  $\psi(x)$ is the set of two Weyl fields with opposite chirality, i.e. $\phi(x)$ and $\chi(x)$. The Dirac equation then reads:
$$
\begin{pmatrix}
0 & \partial_4+i\vec{\sigma}\cdot\nabla\\
\partial_4-i\vec{\sigma}\cdot\nabla & 0
\end{pmatrix}
\begin{pmatrix}
\phi(x)\\
\chi(x)
\end{pmatrix}
=0\ .
$$
Introducing the Dirac gamma-matrices in the chiral basis:
$$
\gamma_k=
\begin{pmatrix}
0 & -i\sigma_k\\
i\sigma_k & 0
\end{pmatrix}\ ,~~~~
\gamma_4=
\begin{pmatrix}
0 & I_2\\
I_2 & 0
\end{pmatrix}\ ,
$$
satisfying the Euclidean Dirac-Clifford algebra:
$$
\{\gamma_{\mu},\gamma_{\nu}\}=\gamma_{\mu}\gamma_{\nu}+\gamma_{\nu}\gamma_{\mu}=2\delta_{\mu\nu}I_4\ ,
$$
the Dirac equation is written in the compact form:
$$
\sum_\mu\gamma_\mu\partial_\mu\psi(x)=0\ .
$$
The lattice theory can be defined by substituting differential operators with finite difference operators. For example:
$$
\partial_\mu\psi(x)\rightarrow\frac1{2a}(\psi(x+ae_\mu)-\psi(x-ae_\mu))\ ,
$$
where $a$ is the lattice spacing and $e_\mu$ is the unit vector along the $\mu$ direction. The corresponding momentum space relation is:
$$
p_\mu\rightarrow\frac ia\sin ap_\mu\ .
$$
However, this correspondence leads to the doubling problem: since $\sin ap_\mu$ is zero at $p_\mu=0$ and $p_\mu=\frac\pi a$ the inverse fermion propagator,
$$
G(p)^{-1}=\frac1a(i\sin ap_4-\sum_k\sigma_k\sin ap_k)\ ,
$$
has 16 zero locations at the 16 corners of the Brillouin zone. Hence, we have a theory describing 16 Weyl fermions as opposed to just one where we started!

\section*{Appendix B: Solution of the secular equation}

In order to compute the zeros of the Weyl operator (\ref{weyl_latt}) one has to solve the secular equation:
\begin{eqnarray*}
0=\det~\W(p)&=&(\cos p_4-1)^2-[\sin p_4+\sum_k(\cos p_k-1)]^2-\sum_k\sin^2p_k\\
&+&2i~(\cos p_4-1)~[\sin p_4+\sum_k(\cos p_k-1)]\ .
\end{eqnarray*}
First notice that the imaginary part vanishes for $p_4=0$, which in turn gives a real part which vanishes only at $\vec{p}=0$. Hence, one can search the other zeros by solving the system:
\begin{eqnarray*}
0&=&(\cos p_4-1)^2-\sum_k\sin^2p_k\\
0&=&\sin p_4+\sum_k(\cos p_k-1)\ .
\end{eqnarray*}
One way to solve this equations is by direct numerical search of zeros in the Brillouin zone. The other method is using the trigonometric identity $\sin^2p_4+\cos^2p_4=1$ which gives the scalar equation:
$$
\sqrt{\left(1-\sqrt{\sum_k\sin^2p_k}\right)^2+\left[\sum_k(\cos p_k-1)\right]^2}-1=0\ .
$$
This reduces the numerical search in three dimensions. In both cases we have checked numerically that there are no more zeros than those stated in the paper. 

However, one can prove that there are no more zeros then the one in the origin and those at $p_k^\pm$. For this let us compute the admissible solutions of the second equation, i.e.
$$
\sum_k(1-\cos p_k)=\sin p_4\leq1~~\Leftrightarrow~~\cos p_1+\cos p_2+\cos p_3\geq2\ .
$$
Since each cosine is bounded by one,
$$
\cos p_1\leq1\ ,~~\cos p_2\leq1\ ,~~\cos p_3\leq1\ ,
$$
one gets three other inequalities:
$$
\cos p_1+\cos p_2\geq1\ ,~~\cos p_2+\cos p_3\geq1\ ,~~\cos p_3+\cos p_1\geq1\ .
$$
Using the same argument one gets three more inequalities on cosines:
$$
\cos p_1\geq0\ ,~~\cos p_2\geq0\ ,~~\cos p_3\geq0\ .
$$
Denoting $x_k\equiv\cos p_k, k=1,2,3$, then the admissible solutions lie within the unit cube of the positive cone above the plane $x_1+x_2+x_3\geq2$. This is the tetrahedron with vertices $(1,1,1),(0,1,1),(1,0,1),(1,1,0)$.

Now we show that there are no solutions in the inner points of the tetrahedron except at the vertices. Since $\cos p_k\geq 0$, we can write:
$$
\cos^2p_k-1=(\cos p_k-1)(\cos p_k+1)\geq\cos p_k-1\ .
$$
One the other hand we have:
$$
\cos^2p_k\leq\cos p_k\ ,~~k=1,2,3\ .
$$
These inequalities, taken together with the previous one, imply that:
$$
\cos^2p_k=\cos p_k\ ,
$$
which is only possible at the vertices of the tetrahedron. The vortex $(1,1,1)$ gives the zero at the origin which is doubly degenerated. The vertices $(0,1,1),(1,0,1),(1,1,0)$ give two zeros each, exactly those located at $p_1^\pm,p_2^\pm$ and $p_3^\pm$. This way, we have given a constructive proof for the zeros of our action.

\end{document}